\documentclass[12pt]{article}
\usepackage{amsmath}
\usepackage{graphicx}
\usepackage{enumerate}
\usepackage{natbib}
\usepackage{url} 

\newcommand{\blind}{1}

\newcommand{\bX}{\boldsymbol{X}}
\newcommand{\bx}{\boldsymbol{x}}
\newcommand{\bY}{\boldsymbol{Y}}
\newcommand{\by}{\boldsymbol{y}}
\newcommand{\bD}{\boldsymbol{D}}
\newcommand{\bd}{\boldsymbol{d}}
\newcommand{\hw}{\hat{w}}

\usepackage{pdflscape}
\usepackage{dcolumn,booktabs}
\newcolumntype{d}[1]{D{.}{.}{#1}}
\usepackage{hyperref}
\usepackage{amssymb}
\usepackage{mathrsfs}
\usepackage{bm}
\usepackage{algorithm}
\usepackage{algpseudocode}
\usepackage{tablefootnote}
\usepackage{appendix}
\usepackage{enumitem}
\usepackage{multirow}
\usepackage{array}
\usepackage{float}
\usepackage{colortbl} 
\usepackage[table]{xcolor} 

\newcommand\mc[1]{\multicolumn{1}{c}{#1}} 

\newcommand{\indep}{\rotatebox[origin=c]{90}{$\models$}}
\newcommand\numberthis{\addtocounter{equation}{1}\tag{\theequation}}
\definecolor{lightgray}{gray}{0.9} 

\begin{document}
\def\spacingset#1{\renewcommand{\baselinestretch}%
{#1}\small\normalsize} \spacingset{1}


\if1\blind
{
  \title{\bf  {\it RAILS}: A Synthetic Sampling Weights for Volunteer-Based National Biobanks -- A Case Study with the {\it All of Us} Research Program}
  \author{Huiding Chen \\
    Department of Biostatistics, Vanderbilt University Medical Center, \\ Nashville, TN, United States \\
    and \\
    Andrew Guide \\
    Department of Biostatistics, Vanderbilt University Medical Center, \\ Nashville, TN, United States \\
    and \\
    Lina Sulieman \\
    Department of Biomedical Informatics, Vanderbilt University \\ Medical Center, Nashville, TN, United States \\
    and \\
    Robert M Cronin \\
    Department of Internal Medicine, The Ohio State University \\ Wexner Medical Center, Columbus, Ohio, United States \\
    and \\
    Thomas Lumley \\
    Department of Statistics, University of Auckland, \\ Auckland, New Zealand \\
    and \\
    Qingxia Chen \thanks{
    corresponding author (cindy.chen@vumc.org)} \\
    Department of Biostatistics, Vanderbilt University Medical Center, \\ Nashville, TN, United States}
  \maketitle
} \fi

\if0\blind
{
  \bigskip
  \bigskip
  \bigskip
  \begin{center}
    {\LARGE\bf  {\it RAILS}: Synthetic Sampling Weights for Volunteer-Based National Biobanks: A Case Study with the {\it All of Us} Research Program}
\end{center}
  \medskip
} \fi

\bigskip
\begin{abstract}
While national biobanks are essential for advancing medical research, their non-probability sampling designs limit their representativeness of the target population. This paper proposes a method that leverages high-quality national surveys to create synthetic sampling weights for non-probabilistic cohort studies, aiming to improve representativeness. Specifically, we focus on deriving more accurate base weights, which enhance calibration by meeting population constraints, and on automating data-supported selection of cross-tabulations for calibration. This approach combines a pseudo-design-based model with a novel Last-In-First-Out criterion, enhancing the accuracy and stability of the estimates. Extensive simulations demonstrate that our method, named \textit{RAILS}, reduces bias, improves efficiency, and strengthens inference compared to existing approaches. We apply the proposed method to the \textit{All of Us} Research Program, using data from the National Health Interview Survey 2020 and the American Community Survey 2022 and comparing prevalence estimates for common phenotypes against national benchmarks. 
The results underscore our method’s ability to effectively reduce selection bias in non-probability samples, offering a valuable tool for enhancing biobank representativeness. Using the developed sampling weights for \textit{All of Us} Research Program, we can estimate the prevalence of the United States population for phenotypes and genotypes not captured by national probability studies. 
\end{abstract}

\noindent%
{\it Keywords:}  Calibration Weighting; Electronic Medical Records; Generalized Raking; Non-Probability Sampling; Prevalence.
\vfill

\newpage
\spacingset{1.9} 
\section{Introduction} \label{sec:intro}

National biobanks, such as UK Biobank (\cite{bahcall2018uk}), the Million Veteran Program (\cite{gaziano2016million}), and the {\it All of Us} Research Program ({\it All of Us}) (\cite{all2019all}), play a pivotal role in medical discovery by providing researchers with access to vast and diverse biological samples, which are essential for understanding diseases, developing new treatments, and advancing personalized medicine. In particular, {\it All of Us} is an ongoing large-scale initiative to collect and study multimodal data from one million or more participants living in the United States (U.S.). Studies have shown significant disparities in disease prevalence when comparing {\it All of Us} data to the broader U.S. population or a representative sample cohort (\cite{karnes2021racial},\cite{chandler2021hypertension}, \cite{aschebrook2022overview}). This could be attributed to the overrepresentation of participants from those traditionally underrepresented in biomedical research in {\it All of Us}. Without appropriate adjustment, the data of {\it All of Us} are not representative of the target population and therefore not generalizable to the U.S. population, which is one of the main challenges of {\it All of Us} (\cite{bianchi2024all}). This limitation arises because
the cohort is volunteer-based \cite{mapes2020diversity}, and data are collected using a non-probability sampling design. 

Unlike non-probability sampling, probability sampling has been regarded as the gold standard in survey statistics and widely adopted in national research and the official census since Neyman's early work in the seminal paper \cite{neyman_two_1934}. The inference of design-based sampling had been well-established in the last century (\cite{hansen_theory_1943}, \cite{berkson_limitations_1946}, \cite{horvitz_generalization_1952}, \cite{kish_confidence_1957}, \cite{cochran_sampling_1977}). Kalton provided a comprehensive overview of the development history (\cite{kalton_developments_2019}). In surveys using probability sampling designs, a standard approach, regardless of initial sampling probabilities, is to further adjust design weights (or base weights). This adjustment aligns weighted survey totals for population subsets defined by specified post-stratification variables with reliable estimates from large national surveys \cite{valliant2013practical}.    

Non-probability sampling methods can be traced back as early as the 1930s, with quota sampling emerging as a prominent example. They were not widely used because the selection bias could be tremendous and the modes of data collection were limited, with a famous example of the failure to predict the 1948 US presidential election (\cite{lusinchi_rhetorical_2017}). However, the rising costs and growing nonresponse rates in large-scale data collection have led to their resurgence. The advent of web-based surveys has further amplified this trend, providing a cost-effective and time-efficient alternative to gather data (\cite{couper2000web}). These methods have gained popularity due to their ability to reach large populations quickly. In response to the growing improper use of non-probability sampling, the American Association for Public Opinion Research (AAPOR) convened a task force to address concerns surrounding these methods \cite{baker2013summary}. The report underlined the importance of post-survey adjustments such as weighting and calibration, although these adjustments cannot fully compensate for the absence of random selection. Ultimately, its use in scientific studies requires careful consideration and rigorous methodological safeguards.

Today, calibration methods that were originally developed to refine probability sampling weights are also applied to non-probability sampling designs with fixed base weights (\cite{valliant2013practical}). This paper seeks to leverage high-quality national surveys to construct synthetic sampling weights for non-probabilistic cohort studies, such as {\it All of Us}. Specifically, we focus on deriving more accurate base weights (or initial weights), which facilitates finding solutions that satisfy calibration constraints, and on automating the selection of cross-tabulations for calibration. Our approach focuses on two types of high-quality national surveys. First, the National Health Interview Survey (NHIS) provides a nationally representative dataset built with a complex probabilistic sampling approach, including validated sampling weights and publicly available individual level data (\cite{botman2000design}, \cite{moriarity2022sample}). This resource enables us to build a design-adapted nested propensity score model (to be introduced in Section~\ref{NPS}) to derive initial base weights and conduct variable selection to identify calibration variables. However, the NHIS sample size, while representative with sampling weight adjustment, is relatively small compared to the entire US population, which can result in undercoverage, particularly for traditionally underrepresented biomedical research groups. To address this limitation, we also incorporate data from the US Census and the American Community Survey (ACS) (\cite{coggins2022understanding}). These larger-scale national surveys provide the resource to construct aggregated (up to lower-order cross-tabulation) but comprehensive demographic information across various geographic levels, broadening demographic coverage and helping to compensate for NHIS limitations.

The remainder of this paper is organized as follows. Section~\ref{method} introduces the proposed method with its detailed algorithm and variance estimate. Section~\ref{simulation} describes the simulation studies conducted to evaluate the performance of the proposed method compared to existing approaches, as well as variations of the proposed method, to identify factors that may influence its performance. In Section~\ref{application}, we apply these methods to the {\it All of Us} data set to estimate the prevalence of the top 15 health conditions in the US population and evaluate these estimates comparing them with results from the Global Burden of Disease Study (GBD) 2019 (\cite{mokdad2018state},\cite{vos2020global}). We also compare the prevalence estimate for factors available in the NHIS but not included in the development of the sampling weights. Finally, we conclude the paper with Section~\ref{discussion}. Deidentified data are available on the researcher workbench of the \textit{All of Us} Research Program located at \url{https://workbench.researchallofus.org}. The codes and scripts for \textit{All of Us} will be made available within the workbench. Related simulation setups and codes are available in \url{https://github.com/\if1\blind{extrasane/RAILS}\fi}

\section{Method} \label{method}

In this section, we will propose a unified workflow that leverages two types of external data resources to construct synthetic sampling weights for the non-probabilistic sample cohorts. 

First, we introduce a method to estimate sampling weights based on nested propensity scores derived from the pseudo-likelihood. Next, we apply forward variable selection to rank variables by their importance for calibration using average likelihood increments in the nested propensity model. Finally, we generate the weights through calibration through raking. In cases where the raking algorithm fails to find a solution that satisfies all constraints, we propose a last-in-first-out strategy to prioritize the algorithm's resolution. We will describe the proposed method in detail below.

\subsection{Setups and notations}
Denote $\bX$ as the vector of auxiliary variables used in the survey sampling, including demographic variables, $\bY$ as the outcome, and $\bD$ as the survey sampling weights. Let $\mathcal{U}$, $\mathcal{U}_{P}$, and $\mathcal{U}_{NP}$ denote the target population, the probability sample cohort, and the non-probability sample cohort with a sample size of $N$, $n_P$, and $n_{NP}$, respectively. We use $P$ or $NP$ in the subscript or superscript to distinguish and highlight probability or non-probability sampling cohorts. 

We assume that both probability and non-probability sample cohorts are proper subsets of the target population: $ \mathcal{U}_{NP} \subsetneq \mathcal{U}, \mathcal{U}_P \subsetneq \mathcal{U} $. In practice, some individuals may be included in both samples through different sources, but this overlap is typically untraceable with deidentified data. Although there are methods to link participant information probabilistically using auxiliary data, such approaches are beyond the scope of this paper. For simplicity, we assume $\mathcal{U}_{NP} \bigcap \mathcal{U}_P = \varnothing$. 
Furthermore, for probability samples, we assume that the survey weights $d_i^P$ are available and adjusted to the target population as $\pi_i^P = 1/d_i^P = Pr(i \in \mathcal{U}_P )$. 

Without loss of generality, we assume the following information is available: 
\begin{align*}
& \mbox{Target population: } \mathscr{F} =\left\{ \mathcal{CO}(\bx_i), i\in \mathcal{U}  \right\} \\
& \mbox{Probability sample cohort: } \mathscr{F}_{P} = \left\{ (\bx_i, \bd_i^P), i\in \mathcal{U}_P  \right\} \\
& \mbox{Non-probability sample cohort: } \mathscr{F}_{NP} = \left\{ (\bx_i, \by_i), i\in \mathcal{U}_{NP}  \right\},
\end{align*}
where $\mathcal{CO}$ represents the coarse information of variables, which can be univariate marginal or lower-order cross-tabulation of multiple variables. 

While the ultimate goal is to estimate the synthetic sampling weights for the non-probability samples, i.e., $\bd_i^{NP}$ for $i \in \mathcal{U}_{NP}$, we focus on the prevalence estimand, i.e., mean($\by_i$) for $i \in \mathcal{U}$, in the simulation and real data applications. This allows us to evaluate the performance of the sampling weight estimates for the non-probability sample cohort.

\subsection{Nested Propensity Score} \label{NPS}

Motivated by \cite{chen_doubly_2020}, we propose the nested propensity score method to leverage $\mathcal{F}_P$ to derive the base weight for $\mathcal{F}_{NP}$.  Consider a binary indicator $R_i$ for all observations in the finite population as $R_i=1$ when $i \in \mathcal{U}_{NP}$, and $R_i=0$ when $i \notin \mathcal{U}_{NP}$.
The propensity score is the conditional probability of a unit assigned to a particular group:
$\pi_i^{NP} = E[R_i \mid \bm{x}_i,y_i] = Pr(R_i \mid \bm{x}_i,y_i)$

To utilize auxiliary information from the probability samples, the same assumptions for \textit{strong ignorability} are considered (\cite{rosenbaum_central_1983},\cite{rubin1976inference}):
\begin{enumerate}\label{strongignor}
    \item \textit{Conditional Independence}: The response variables are independent of the assignment given the set of auxiliary variables $(y^{NP}_i,y^P_i) \indep R_i | \bm{x}_i$
    \item \textit{Positivity}: For every possible combination of $\bm{x}$, there is a nonzero probability of being assigned to both designs
    $Pr(R_i|X = \bm{x}) \in (0,1)$
\end{enumerate}
Then the propensity scores are
$\pi_i^{NP} =  Pr(R_i \mid \bm{x}_i,y_i) = Pr(R_i \mid \bm{x}_i)$.
The \textit{strong ignorability} is essential to derive the synthetic sampling weights, as no information is needed in the response when modeling propensity scores. 

Note that fitting a standard classification model to the combined non-probability and probability samples, $\mathcal{F}_{NP} \bigcup \mathcal{F}_P$, generally results in a biased estimate of the propensity score (\cite{Rivers2007SamplingFW},\cite{valliant_estimating_2011}). 
Assuming $\bm{\pi}_i^{NP} = m(\bm{x}_i,\bm{\theta}_i)$, the log-likelihood can be written as:
\begin{align*}
    \ell(\bm{\theta}) &= \sum_{i \in \mathcal{U}}\big[ R_i \log(\bm{\pi}_i^{NP}) + (1-R_i)\log(1 - \bm{\pi}_i^{NP}) \big] \\
    & = \sum_{i \in \mathcal{U}_{NP}} \log \left(\frac{\bm{\pi}_i^{NP}}{1- \bm{\pi}_i^{NP}}\right) +  \sum_{i \in \mathcal{U}} \log(1 - \bm{\pi}_i^{NP})  \numberthis\label{1}
\end{align*}
Notice that the second term in \eqref{1} cannot be directly used since not all subjects are included. Therefore, we implemented the Horvitz-Thompson plug-in estimators using probability samples. Thus the pseudo-likelihood is specified as \cite{chen_doubly_2020}:
\begin{align*}
    \ell^*(\bm{\theta}) &=  \sum_{i \in \mathcal{U}_{NP}} \log \left(\frac{\bm{\pi}_i^{NP}}{1- \bm{\pi}_i^{NP}}\right) +  \sum_{i \in \mathcal{U}_P} d_i^P \log(1 - \bm{\pi}_i^{NP}) 
\end{align*}

Considering the binary form of the indicator and its concise form, the natural choice would be logistic modeling $\bm{\pi}_i^{NP}=m(\bm{x}_i,\bm{\theta}) = \left[1 + exp(-\bm{x}_i^{\intercal} \bm{\theta})\right]^{-1}$
\begin{equation}
\ell^*(\bm{\theta}) =  \sum_{i \in \mathcal{U}_{NP}}\bm{x}_i^{\intercal} \bm{\theta}  -  \sum_{i \in \mathcal{U}_P} d_i^P \log\big[ 1 + exp(\bm{x}_i^{\intercal} \bm{\theta})  \big]
\label{nps-lik}
\end{equation}

As the likelihood in (\ref{nps-lik}) is different from the likelihood function from the weighted logistic regression model, we use the Newton-Raphson method to find $\bm{\theta}_{MLE}$ via iteration as follows:
\begin{equation}
\bm{\theta}^{(i+1)} = \bm{\theta}^{(i)} - [H(\bm{\theta}^{(i)})]^{-1}S(\bm{\theta}^{(i)}), \label{delta:theta} 
\end{equation} 
\begin{equation}
\mbox{where } S(\bm{\theta}) = \sum_{i \in \mathcal{U}_{NP}}\bm{x}_i  -  \sum_{i \in \mathcal{U}_P} d_i^P \bm{\pi}_i^{NP} \bm{x}_i,
    H(\bm{\theta}) =  - \sum_{i \in \mathcal{U}_P} d_i^P m(\bm{x}_i,\bm{\theta}) [1 - m(\bm{x}_i,\bm{\theta})] \bm{x}_i \bm{x}_i^{\intercal} 
  \label{hmat}
\end{equation}
The resulting $\widehat{\bd}_i^{NP}=1/\widehat{\bm{\pi}}_i^{NP}=1 + exp(-\bm{x}_i^{\intercal} \widehat{\bm{\theta}})$ serves as the initial base weight for the non-probability sampling cohort. The procedure can be illustrated as follows:

\begin{algorithm}[ht]
\caption{Nested Propensity Score for Non-probability Sampling Weights}\label{alg:nps}
\begin{algorithmic}[1]
\Procedure {NPS}{$S_c,X^{NP},X^{P},\bm{d}^{P}$} 
    \Repeat{ Compute iterative residuals $\Delta{\bm{\theta}^{(i)}}$ } \Comment{See equation (\ref{delta:theta})}
        \Until{$\left\lVert\Delta{\bm{\theta}^{(i)}}\right\lVert_2 \leq \epsilon_1$} 
\EndProcedure \Comment{Output: $\widehat{\bm{d}}^{NP},\ell_c^{*}$}
\end{algorithmic}
\end{algorithm}

\subsection{Generalized Raking} \label{rake}
With the sampling weights $\widehat{\bd}_i^{NP}$,
the total of outcome can be estimated as $\hat{y}_{tot} = \sum_{i \in \mathcal{U}_{NP}} \widehat{d}_i^{NP} y_i$. However, this estimation is generally biased or unstable, especially when the propensity model is misspecified, the relative size $n_P/n_{NP}$ is small, or the conditional independence of the outcome is not perfectly satisfied (\cite{meng2018statistical}). We implement generalized raking to improve $\widehat{d}_i^{NP}$, so that the weighted auxiliary variables are aligned to the known population totals. We denote the sampling weight of the raking as $w_i^{NP}$. 

The problem can be stated as an optimization problem with constraints (\cite{deville_generalized_1993}).
Let $G(\cdot)$ be a distance function of $w_i^{NP}/\widehat{d}_i^{NP}$,  \textrm{s.t.} $G(1) = 0$, $G'(1) = g(1)=0$, and $G''(1)=1$. The algorithm can be stated as:
\begin{align*}
\min_{w} &\quad D(w) = \sum_{i \in\mathcal{U}_{NP}} \widehat{d}_i^{NP} G\bigg(\frac{w_i^{NP}}{\widehat{d}_i^{NP}}\bigg) \\
\textrm{s.t.} &\quad \sum_{i \in \mathcal{U}_{NP}} w_i^{NP} \bm{x}_i = \bm{T}^{pop}, \sum_{i \in \mathcal{U}_{NP}} w_i^{NP} = \sum_{i \in \mathcal{U}_{NP}} \widehat{d}_i^{NP}= N, 
\mbox { and } w_i^{NP} \geq 0 \numberthis\label{alg:raking}
\end{align*}
The first constraint is not limited to the univariate margins. The interaction terms could also be raked with the corresponding population grouping variables. Unlike post-stratification, raking allows multiple covariates to be calibrated without constructing a complete cross-classification (\cite{lumley_complex_2010}).

The optimization problem can be solved using the Lagrange multiplier $\bm{\lambda}$. The Lagrange multiplier is constrained by: 
\begin{equation*}
w_i^{NP} = \widehat{d}_i^{NP} F(\bm{x}_i^{\intercal}\bm{\lambda}), 
\mbox {  where  }
F(\cdot) = g^{-1}({\cdot}) 
\end{equation*}
The problem is solved iteratively with two steps: (a) Given a specific distance function $G(\cdot)$ and its derivative $g(\cdot)$, solve the constraint \eqref{alg:raking} for $\bm{\lambda}$; (b) Determine the calibrated weights $w_i^{NP}$ by constraint conditions for the $\bm{\lambda}$ found in step (a).

By selecting different distance functions, the solution leads to different calibration estimators. The linear method $G(x) = (x-1)^2/2$ corresponds to \textit{generalized regression estimator} (GREG) for the responsive variable. In the proposed method, the \textit{multiplicative method} $G(x) = x\log x - x + 1$ is applied, which leads to the raking. We also implement iterative proportional fitting (IPF) to derive $\bm{\lambda}$ at a given tolerance. The following pseudo-algorithm describes the idea of generalized raking, and the resulting $d_i^{NP,(i)}$ at convergence is $\hat{w}_i^{NP}$.
\begin{algorithm}[ht]
\caption{Generalized Raking}\label{alg:rake}
\begin{algorithmic}[1]
\Procedure{Calibration Raking}{$S_c,\widehat{\bm{d}}^{NP},\bm{X}^{NP},\bm{T}_{pop}$} 
    \Repeat \Comment{See~\ref{alg:raking}}
        \For {$j = 1,\cdots,k, k =|S_C|$} 
        \State $\widehat{\bm{d}}^{NP,(i)} =\left[ {T_j}^{pop} / \sum_{i \in \mathcal{U}_{NP}} \widehat{d}^{NP,(i-1)}x_{ij} \right] \cdot  \widehat{\bm{d}}^{NP,(i-1)}$
        \State $i = i + 1$
        \EndFor
    
    \Until{$ max\left\lVert \bm{T}^{pop} - \sum_{i \in\mathcal{U}_{NP}} \widehat{\bm{d}}^{NP,(i)}\bm{x}_{i} 
    \right\lVert \leq \epsilon_1 \lor i = n_{max}$} 
\EndProcedure
\end{algorithmic}
\end{algorithm}

\subsection{Variable Selection}

To fully utilize the information from the non-probability samples with auxiliary covariables, it is intuitive to include higher-order interactions of covariates into the nested propensity score model and/or the calibration procedure. However, including higher-order interactions could result in the singularity for the inverse of the Hessian matrix in \eqref{hmat} or the rank deficiency in the design matrix for calibration. Moreover, subgroups with zero or few counts could lead to extreme weights or non-convergence. Although truncation or setting bounds on the estimated weights could mitigate the problem, finding the reasonable or optimal cutoff point in practice is rather formidable. We proposed a forward variable selection scheme to select the higher-order interactions to the working model. It is a greedy algorithm that maximizes pseudo-likelihood when adding new variables to the current model.

The set of all available variables to include, consisting of univariate and higher-order interactions, is denoted as $S = \{x_1,x_2,\cdots,x_1:x_2,x_1:x_3,\cdots\} = \{z_1,z_2,\cdots\}$, and the set for the contemporary working model as $S_w$. The pool of variables to be selected is $S_c = S\backslash S_w = S_w^{C}$. The index set of S is $I = \{1,2,\cdots,J\}$, and $I_s$ for the working model, $I_c$ for the pool. 

Given $S_w$, we compute the increments $\Delta\bm{\ell} = \bm{\ell}_{alt} - \bm{\ell}_{null}$ in the log pseudo-likelihood in the propensity score model when adding each element from $S_c$, then the p-values $\bm{p}_c$ of the likelihood ratio test are calculated based on the increment and the degree of freedom $\bm{\nu}$ of that element. By Wilk's Theorem,
$p_j = Pr(\chi^2_\nu\geq 2\Delta\ell_j).$

Although multiple p-values could be significant, we update the set for the working model by adding the most influential element with a significant p-value and a maximum average increase $\Delta\ell_j / \nu_{j}$. The algorithm can be illustrated in Algorithm~\ref{alg:vs}.

\begin{algorithm}[htp]
\caption{Greedy Variable Selection}\label{alg:vs}
\begin{algorithmic}[1]
\Procedure{GVS}{$S_c,S_w,\ell_c^{*}$} 
    \Repeat{ 
        \State Compute $(\Delta\ell_j, p_j)$ for every element in $S_c$
        \If{$\exists j\in S_c, p_j < \alpha$}
            \State $A = \{k\in I_c:p_k <\alpha\}$
            \State $K = \mathop{\arg\max}\limits_{k} \{\Delta\ell_k / \nu_{k}\}, k\in A$ 
            \State $S_c \leftarrow S_c/\{z_K\}$  
            \State $S_w \leftarrow S_w\bigcup\{z_K\}$
            \State $\ell_c^{*} \leftarrow \text{NPS}(S_c,X^{NP},X^{P},\bm{d}^{P})$
        \Else \State \textbf{Break}
        \EndIf
        }
    \Until{$S_c = \varnothing$} 
\EndProcedure \Comment{Output: $S_w$}
\end{algorithmic}
\end{algorithm}

Ideally, the selected variables are more likely to have a successful calibration as in Algorithm~\ref{alg:rake}. However, if the calibration fails to produce a set of weights, the Last-In-First-Out (LIFO) strategy is used. To perform LIFO, we treat the variables in the initial model as a stack, and the last selected variable will be first removed from $S_w$, followed by regenerated weights as in Section~\ref{NPS} and re-raking as in Section~\ref{rake}. The procedure continues until the calibration convergence is reached or no variables are left in $S_w^{*}/S_c$. We refer to our proposed algorithm as a \underline{R}aking with \underline{A}ssisted Nested Propens\underline{i}ty Score and \underline{L}IFO \underline{S}election  (\textit{RAILS}).

\vspace{-0.1in}
\subsection{Variance Estimate}
The prevalence of the outcome variable can be estimated by
\begin{equation}
 \hat{\mu}_y={\sum_{i \in \mathcal{U}_{NP}} \hw_i y_i} \Big/{\sum_{i \in \mathcal{U}_{NP}} 
 \hw_i}
 \label{est}
\end{equation}
where $\hw_i$ is the sampling weight at convergence in Algorithm \textit{RAILS} described above.

The estimate in (\ref{est}) is known as the estimator of $\pi\text{-}$ for the mean ratio. The numerator stands for the weighted sum of observed samples, and the denominator is the sum of weights if the true size of the target population is not known.
 We adopt the variance for $\hat{R}$ approximated with Taylor linearization. More details for derivation can be found in section 5.8 of S{\"a}rndal, 2003 (\cite{sarndal2003model}):
\begin{align*}
    \hat{R}  & = \frac{\sum_{i\in \mathcal{U}} w_i y_iI_i^{NP}}{\sum_{i\in \mathcal{U}} w_i I_i^{NP}} \approx R + \frac{1}{N} \sum_{i\in \mathcal{U}} I_i^{NP}\left( \frac{y_i}{\pi_i} - \frac{R}{\pi_i} \right) \\ 
 \hat{V}(\hat{R}) & = \frac{1}{\hat{N}^2} \sum_{i=1}^n \sum_{j=1}^n I_i^{NP} \Delta_{ij} \left( \frac{y_i - \bar{y}_w}{\pi_i} \right) \left( \frac{y_j - \bar{y}_w}{\pi_j} \right) 
 \end{align*} 
\begin{equation*}
\mbox{where } 
    E[\hat{R} - R] \rightarrow 0,
    \Delta_{ij} = Cov(I_i^{NP},I_j^{NP}) = \begin{cases}
    \pi_{ij} - \pi_{i}\pi_{j}, \mbox{ if } i \neq j\\
    1 - \pi_
    {i}, \mbox{ if } i = j\\
    \end{cases}
\end{equation*}
Since the inclusion of different individuals is independent (given their inclusion probabilities),  the covariance $\Delta_{ij}$ is nonzero only when $i=j$.
\begin{align*}
    \hat{V}(\bar{y}_w) = \frac{1}{\left( \sum_{i\in \mathcal{U}_{NP}} w_i \right)^2} \sum_{i\in \mathcal{U}_{NP}} w_i^2 (y_i - \bar{y}_w)^2
    \approx \frac{1}{\hat{N}^2} \sum_{i\in \mathcal{U}_{NP}} w_i^2 (y_i - \bar{y}_w)^2
\end{align*}
The second approximation is equality if the population size is known to be equal to the sum of weights. These conditions will hold for calibration estimators where the estimation has converged.

\section{Simulation}\label{simulation}
\subsection{Set-up}
In our simulation, the relative sample sizes of the finite population and sample cohorts are chosen to mimic the motivating study so that $N = 3,340,000$, $n_{NP} = 35,000$, and $n_P = 5,500$. We simulate the auxiliary variables $\bm{x}_j, j\in\{1,\cdots,5\}$ to represent essential covariates for both demographics and socio-economic status. In particular, $x_1$ is simulated from $\mathcal{N}(20,25)$ and categorized as ordinal categorical at the observed; $x_2$, $x_3$, $x_4$, and $x_5$ are simulated from $Bernoulli(0.65)$, truncated Pareto with parameter $\gamma = 8,\nu = 300,k = 1$, multinominal distribution with probabilities $ \bm{p} =(0.277,0.287,0.431,0.005)$, and $Bernoulli(0.1\cdot x4)$, respectively. The auxiliary variables are classified on their quantiles as in the motivating study, though they could have been originally continuous, such as age or household income. In addition to satisfying \textit{strong ignorability}, the selection probability given the auxiliary variables is independent between units. The outcome, selection of non-probabilistic cohort, and selection of probabilistic cohort are sampled from Bernoulli distribution $y_i \sim Bernoulli(p_i)$, $I_i^{NP} \sim Bernoulli(\pi_i^{NP})$, and $I_i^P \sim Bernoulli(\pi_i^P)$ using logistic regression models with up to two-way interactions. 

The variables $x_4$ and $x_5$ are considered the add-on variables that may influence a volunteer’s decision to participate (that is, $\pi_i^{NP}$). These variables may or may not be included in the sampling probability for the probabilistic cohort ($\pi_i^{P}$) or affect the prevalence of the outcome ($p_i$). To reflect the complexity of non-probability samples, the distributions of $x_4$ and $x_5$ are more imbalanced, representing motivations such as health reasons for participation. In contrast, variables $x_1$, $x_2$, and $x_3$ are core auxiliary variables that affect all three models: $(p_i)$, $\pi_i^{NP}$, and $\pi_i^{P}$. We consider different models for $p_i$ and $\pi_i^P$ to evaluate the impact of auxiliary variables by altering the interaction effects and the add-on variable effects in the response variable, and whether the survey weights $d_i^P$ in probability samples account for the add-on variables. For $\mathcal{U}_P$, we assume that the survey weights $d_i^P$ are accurate, such that $d_i^P$ are the inverse of the true sampling probability.

The following models were considered for $p_i$, $\pi_i^{NP}$, and $\pi_i^P$:
\begin{align}
    \mathrm{logit}(p_i) & = \alpha_0 + \alpha_1 x_{i1} +\alpha_2 x_{i2} + \sum\nolimits_{j=1}^3 \alpha_{2+j} I(x_{i3}=j) + \alpha_6 x_{i1}x_{i2} + \alpha_7 x_{i2}x_{i3} \nonumber \\
    &  + \sum\nolimits_{j=1}^3 \alpha_{7+j} I(x_{i4}=j) + \alpha_{11} x_{i5} \\
   \mathrm{logit}(\pi_i^{NP}) &= \beta_0 + \beta_1 x_{i1} +\beta_2 x_{i2} + \sum\nolimits_{j=1}^3 \beta_{2+j} I(x_{i3}=j) + \beta_6 x_{i1}x_{i2} + \beta_7 x_{i2} x_{i3} \nonumber \\
   & + \sum\nolimits_{j=1}^3 \beta_{7+j} I(x_{i4}=j) + \beta_{11} x_{i5} \\
   \mathrm{logit}(\pi_i^P) &= \gamma_0 + \gamma_1 x_{i1} +\gamma_2 x_{i2} + \sum\nolimits_{j=1}^3 \gamma_{2+j} I(x_{i3}=j)+\gamma_6 x_{i1}x_{i3} + \gamma_7 x_{i2}I(x_{i3}>0) \nonumber \\
   &  + \sum\nolimits_{j=1}^3 \gamma_{7+j} I(x_{i4}=j) + \gamma_{11} x_{i5}
\end{align}

By varying the values of the parameters, we consider the following four scenarios.
\begin{itemize}\label{scenario}
 \item[(S1)] Scenario 1: $\alpha_j=\gamma_j=0$, where $j=6,...,11$. In this scenario, there are no interaction effects among the core auxiliary variables in the outcome model ($p_i$) or the selection model for the probability sample ($\pi_i^P$), and the add-on variables for the non-probability sample also have no effect on these models. Under these conditions, raking based on univariate margins is expected to perform well.
 \item[(S2)] Scenario 2: $\alpha_j=\gamma_j=0$ when $j=6,7$, and $\alpha_j \neq 0$ and $\gamma_j \neq 0$ when $j=8,..,11$. In this scenario, no interaction effects are present among the core auxiliary variables in either the outcome model or the probability sample. 
 The effects from the add-on variables are present in both models.
 \item[(S3)] Scenario 3: $\alpha_6 \neq 0$, $\alpha_7 \neq 0$, $\alpha_k=0$, $\gamma_j=0$, where $k=8,...,11$ and $j=6,...,11$. In this scenario, there are no interaction effects among the core auxiliary variables in the outcome model $p_i$; however, there are interactions in the selection model for the probability sample $\pi_i^P$. The add-on variables have no effect on either model. 
 \item[(S4)] Scenario 4: $\alpha_j \neq 0$, $\gamma_j=0$, where $j=6,...,11$. In this scenario, there are no interaction effects among the core auxiliary variables, and the add-on variables have no effect in the outcome model $p_i$. However, both interaction effects and add-on variable effects are present in the model for $\pi_i^{P}$. 
 \item[(S5)] Scenario 5: $\alpha_j \neq 0$, $\gamma_j \neq 0$, where $j=6,...,11$. In this scenario, interaction effects among the core auxiliary variables, as well as effects from the add-on variables, are present in both the models for $p_i$ and $\pi_i^{P}$.
 \end{itemize}
 We expect the proposed \textit{RAILS} method to improve performance by raking at meaningful cross-tabulation margins except for Scenario S1.

\subsection{Evaluated Methods} \label{EvalMethods}

To examine the performance of the weights from different methods on non-probability samples, the following estimators are considered: 
\begin{enumerate}[label=(\roman*)]
    \item \textit{naive}: 
    This method estimated the prevalence as the unweighted mean of the response variable in the non-probability samples in $\mathcal{U}_{NP}$.
    \item \textit{oracle}: 
    The weighted mean of $y$ was calculated using the true sampling probabilities to generate the non-probability samples. This allowed for the direct application of design-based inference. However, this information was unknown in practice and thus included only as a benchmark for comparison.
    \item \textit{cal-1}:
    This method constructed weights through raking using the univariate margins in $\mathcal{U}$, starting with equal initial weights. 
    \item \textit{nps-1}: This is an inverse probability weighting (IPW) method, with weights calculated via nested propensity score that includes only main effects.
    \item \textit{nps-cal-1}:
    This is a hybrid method, with weights calculated via nested propensity score including main effects followed by raking with univariate margins in $\mathcal{U}$.
    \item \textit{vs-nps}:
    This is an IPW method with variable selection. After selecting the variables, the nested propensity model is updated and the weights are recalculated.
    \item \textit{vs-rake}:
    This is a hybrid method where the weights are calculated with the workflow introduced as in section~\ref{method}. The initial raking variable list $S_W$ includes all the univariate variables and $S_c$ of all two-way interactions.
    \item \textit{RAILS}:
    This is the proposed nested propensity score and variable selection enhanced raking method. If the calibration raking fails to converge with $S_w$, the working model will rule out elements in a stack-like fashion --- Last-In-First-Out --- and calibration is applied. The procedure is stopped until the raking is successful or no element is left in $S_w$.
\end{enumerate}    

 Some other estimators were considered but not listed due to significant drawbacks.

\begin{enumerate}
    \item Calibration with the linear method. This method is generally faster and less biased. However, it can produce negative weights. These are mathematically valid, but not accepted by some analysis software, and can cause interpretation difficulties in subpopulations. Official statistics agencies often choose methods such as raking to avoid negative weights.
    \item \textit{cal-2}, 
    and \textit{nps-2}:
    The propensity or calibration model contains the main effects and all two-way interactions. To guarantee the design matrix is full-rank, the pivoting strategy via QR-Decomposition is implemented. The collinear interactions with the lowest ranking in pivots will be eliminated to make a fair comparison to the greedy variable selection. However, these two methods hardly converge when higher orders are involved, since they require the design matrix to be nonsingular and the calibration equation to be nondivergent. 
\end{enumerate}

\subsection{Results}

\begin{figure}
\includegraphics[width=1\linewidth]{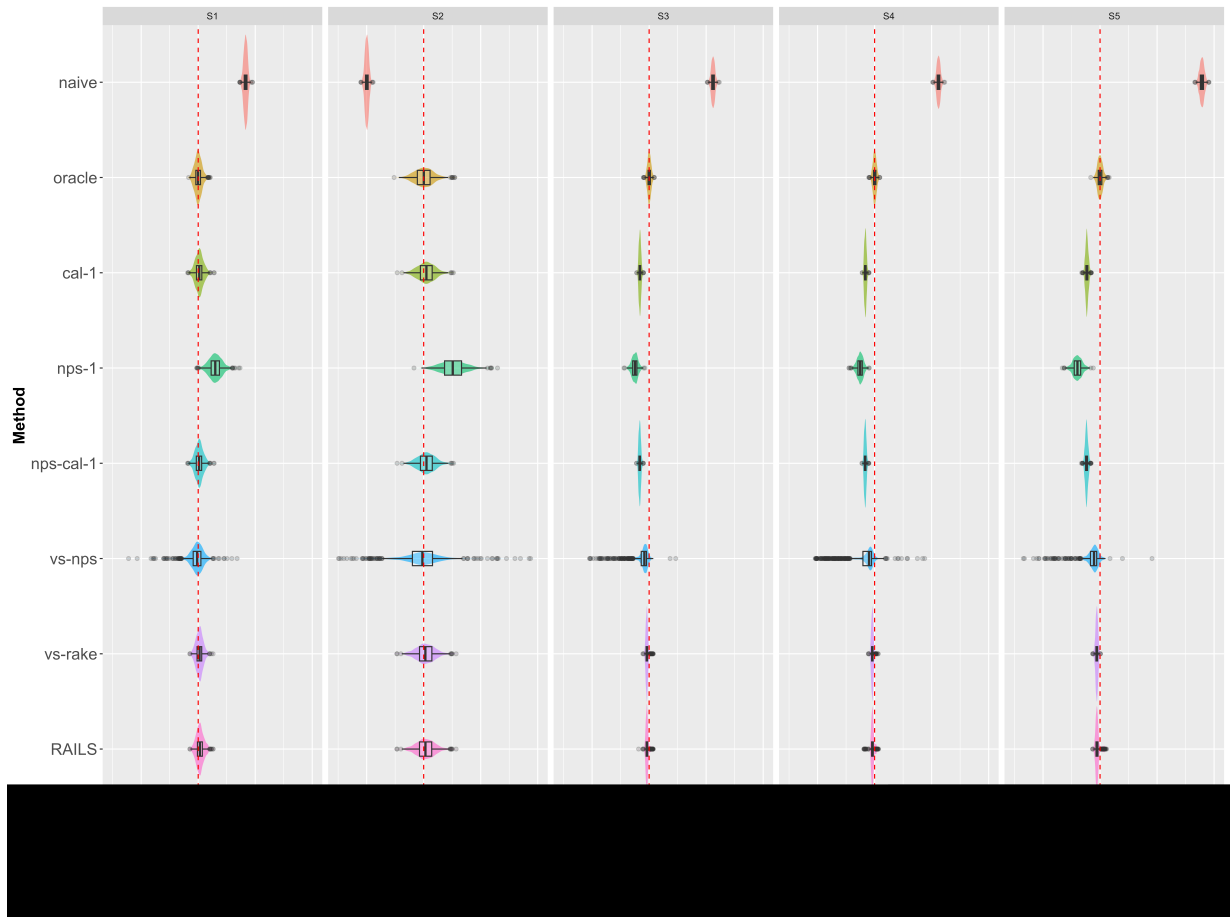}
    \caption{Boxplots for the bias of the estimated population mean for methods in different scenarios. The methods and set-ups are introduced in Sections~\ref{scenario}  and~\ref{EvalMethods}. }
\label{fig:sim}    
\end{figure}

In the four scenarios, estimators were evaluated based on relative bias, coverage, and convergence. In addition to nominal coverage, we calculated oracle coverage - assuming the average bias for each method is known through an oracle - to help identify potential undercoverage resulting from bias. The results, based on 1000 simulations, are summarized in Tables~\ref{tab:sim1} and ~\ref{tab:sim2} with the bias distributions presented in Figure~\ref{fig:sim}.

The simulation results reveal several insights. In Scenario S1, where there are no interaction effects or add-on variable effects on the outcome or the selection of the probability sample, the univariate raking method (\textit{cal-1}) shows optimal performance, achieving minimal bias and accurate coverage probabilities. However, \textit{cal-1} lacks robustness in more complex scenarios. Its bias is substantial enough that, despite yielding over 98\% coverage probability (CP) for the debiased oracle CP, its nominal CPs drop to zero in Scenarios 3 through 5. The hybrid method (\textit{nps-cal-1}) performs similarly to \textit{cal-1}, suggesting that enhancing the base weights has a limited effect on overall performance. However, in the absence of suitable initial weights, calibration raking with high-order variables (\textit{cal-2}) encounters matrix singularity and fails to converge in over 99\% of the simulations. 

When used in isolation, IPW estimators suffer from significant biases and severe under-coverage. Calibration is crucial to mitigate these biases. Our proposed \textit{RAILS} method, which combines propensity score weighting with calibration raking, addresses this issue. By incorporating the information from the probability sample, our method improves the raking even when the sampling probability model ($\pi_i^P$) does not account for the add-on variables. The \textit{vs-rake} method, while performing similarly to \textit{RAILS} when it converges, fails to find a solution that satisfies the calibration constraints in 6\%-27.5\% of simulations.
The LIFO strategy proves to be a robust and automated approach to identifying feasible raking constraints. The \textit{RAILS} estimator performs exceptionally well when interaction effects are important in the response variable. Under Scenario S1, its performance remains acceptable, even though simple calibration methods perform slightly better. In more complex scenarios, particularly S3 to S5, models incorporating two-way interactions would likely outperform others if calibration succeeds. Our method guarantees convergence for raking while efficiently leveraging information from probability samples.

\section{Application on the {\it All of Us} Research Program}\label{application}

In {\it All of Us}, participants aged 18 years and older are enrolled in healthcare care provider organizations or community partners after informed consent (\cite{all2019all}). Our objective is to develop synthetic sampling weights for participants who completed the baseline survey and provided EHR data in the {\it All of Us} Researcher Workbench V7 Controlled Tier release. The program focuses on historically underrepresented populations in biomedical research. 
To support this focus, we calibrate key demographic factors based on the definition in {\it All of Us} (\cite{all2019all}), including age, sex, race/ethnicity, region, household income, education attainment, and house tenure, as well as their cross-tabulations up to three-way, with 1240 levels in all.

For this analysis, we exclude 73,492 participants lacking basic demographic information, resulting in a final sample of 213,520 participants. The NHIS study from the Centers for Disease Control and Prevention (CDC) in 2020 is chosen as the external probability sample. NHIS is design-based with survey weights available and consists of essential characteristics about the health of civilians. Under the same procedure to demand nonmissing covariates, we include 24,678 samples from the NHIS study. As for the national totals, the Public Use Microdata Sample (PUMS) Data from the Census Bureau’s ACS 2022 study serves as our gold standard of population margins; the number of adults settles at 261,048,645. We use the sampling weights in PUMS 2022 to create the cross-tabulation margins for the US population.

We derive the weights using the \textit{naive}, \textit{cal-1}, \textit{nps-1}, \textit{ns-cal-1}, \textit{nps-2}, and \textit{RAILS} methods, as described in Section~\ref{EvalMethods}. To evaluate the impact of calibrating on higher-order interactions, we also include \textit{nps-cal-2}, which is a hybrid method in which weights are calculated via a nested propensity score with main effects and two-way interactions, followed by raking with univariate and two-way interaction margins in $\mathcal{U}$. The final \textit{RAILS} method calibrated on all one-way and two-way interactions of the key demographic factors defined above, as well as the three-way interaction between sex, race/ethnicity, and house tenure.

\subsection{Compare Disease Prevelance Estimates with Global Burden Disease Study} \label{Sec:AoU1}

To assess the impact of these weighting methods on \textit{All of Us}, we estimate the prevalence of the top fifteen health conditions in the US population, using the six weighting methods, and compare them to the disease prevalence reported in GBD 2019 (\cite{mokdad2018state}, \cite{vos2020global}, \cite{bianchi2024all}). Disease status is defined using ICD codes from the EHR data, following the method used by (\cite{zeng2024comparison}). Consistent with the literature on phecode-based phenotyping, individuals with at least two documented phecodes in the EHR are classified as cases for each phenotype. 
Note that although {\it All of Us} was launched nationally in 2018, the EHR data include the available historical health records before launching for consented participants.

\begin{figure}
    \includegraphics[width=1\linewidth]{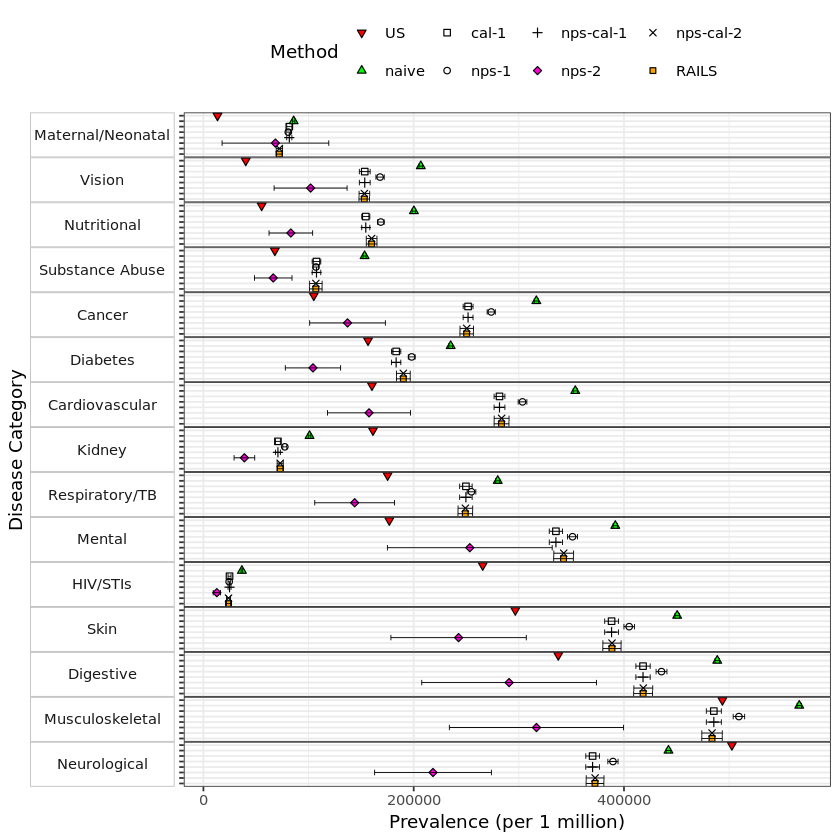}
    \caption{The weighted and unweighted estimates from the {\it All of Us} research program against those in the U.S. general populations estimated by the Global Burden of Disease Study (GBD) 2019 (\cite{mokdad2018state},\cite{vos2020global}) for the top health conditions in the U.S. population (\cite{bianchi2024all}). Method \textit{US} is the general U.S. prevalence estimate from GBD 2019 study; methods \textit{naive}, \textit{cal-1}, \textit{nps-1}, \textit{ns-cal-1}, \textit{nps-2}, and \textit{RAILS} were described in section~\ref{EvalMethods}; and method \textit{nps-cal-2} was described in section~\ref{application}.}
    \label{fig:AoUFull1}
\end{figure}

Table~\ref{tab:AoU1} compares the auxiliary variables between the proportions of the population in PUMS 2022 and the study proportions from \textit{All of Us}, both unweighted and weighted using the \textit{RAILS} method, showing that the weighted proportions align closely with the population values. The prevalence of the disease is estimated using synthetic sampling weights derived from each method, as illustrated in Figure~\ref{fig:AoUFull1}. Although some discrepancies remain, the weighted disease prevalence estimates are generally closer to US disease prevalence benchmarks than the unweighted estimates for most conditions. The \textit{nps-2} method produces estimates that deviate significantly from other weighted estimates, exhibiting much wider confidence intervals due to extreme weights. This issue likely arises from undercoverage associated with the relatively small sample size of NHIS. In our simulation studies, IPW with higher-order interactions, without the additional raking step, often resulted in substantial bias and high variability, making it unreliable. The \textit{nps-2} method is not included in Tables~\ref{tab:sim1} and ~\ref{tab:sim2} due to non-convergence in most simulations. Method \textit{vs-nps}, an improved version of \textit{nps-2}, still shows large bias and variability. The emaining discrepancies observed in Table~\ref{tab:AoU1} could be attributed to several factors, including differences in ICD coding, cohort composition, unmeasured sampling variables, and temporal changes in prevalence. 

\subsection{Compare Health Outcome Estimates with NHIS} \label{Sec:AoU2}

To further evaluate the performance of the developed sampling weights, we compared the weighted and unweighted prevalence estimates for health survey questions included in both NHIS and \textit{All of Us} but not used in the development of the sampling weights. These factors include common visit care place (a minimally dynamic outcome), self-reported health (a moderately dynamic outcome), and insurance coverage (the most dynamic outcome).

A new challenge arises due to missing data for these factors. We consider two methods to address this issue:
\begin{itemize}
\item Recalibrate: This method focuses on the cohort with complete information for the studied outcome. Although this provides a straightforward implementation of the proposed method, it implies that external datasets would need to be provided to practitioners, and they would have to implement the method for their specific cohorts, which may limit its broader applicability.
\item Double-weighting approach: This more practical method incorporates a second set of weights to model the probability of a participant being in the complete cohort given their inclusion in the \textit{All of Us} cohort. Specifically, we model the missing data mechanism for each outcome using a logistic regression model that includes all demographic and social-status variables for \textit{RAILS}. The resulting missing-data weight is derived as the inverse of the estimated probability of not missing. The final weight is obtained by multiplying the \textit{RAILS} weights by the missing-data weights, followed by normalization to align with the size of the US population.
\end{itemize}
As shown in Table~\ref{tab:AoU2}, the double-weighting estimates from \textit{All of Us} reduced the total absolute discrepancy by 44.76\% (3.95\% weighted vs 7.15\% unweighted) for the least dynamic outcome, common visit care place. For the moderately dynamic outcome, self-reported health, the discrepancy was reduced by 28.16\% (24.80\% weighted vs 34.52\% unweighted). However, for the most dynamic outcome, insurance coverage, the discrepancy increased by 24.25\% (12.5\% weighted vs. 10.06\% unweighted). The recalibrate method yields similar estimates with the same trend as the double-weighting, supporting the usage of the double-weighting method in practice in the presence of missing observations for the studied outcome. These results indicate that weighted estimates improve the prevalence estimate for stable outcomes while highlighting the challenges of more dynamic outcomes. 

It is also worth mentioning that differences could also be attributed to survey mode, study design, and survey answer options, among others. For example, NHIS 2020 primarily used a telephone interview mode due to the COVID-19 pandemic, with a shift from in-person interviews to phone interviews starting in March 2020 (\cite{NHIS2020}), while participants of \textit{All of Us} completed surveys online; NHIS's data are cross-sectional, while for \textit{All of Us}, participants can complete the available surveys at any time after enrollment with the dates completed the survey spread over years since the program initiated in 2017. As for answers, for common visit care place, NHIS's answer options include ``a doctor's office or health center" and ``urgent care center or clinic in a drug store or grocery store," while \textit{All of Us}'s corresponding options are ``doctor's office" and ``urgent care."

\section{Discussions}
\label{discussion}

The lack of representativeness in non-probability sampling designs poses a significant challenge, particularly as costs and non-response rates continue to rise. In discussing the balance between data quality and quantity, Meng (2018) emphasized the importance of prioritizing data quality when working with large datasets, suggesting that smaller but higher quality datasets should be weighted more heavily than size alone would dictate when combining multiple data sources for population inference (\cite{meng2018statistical}). However, effectively leveraging a smaller, high-quality probability sampling cohort alongside a larger, lower-quality non-probability sampling cohort remains a challenge, particularly when the goal is to develop sampling weights for the non-probability cohort. 

Across the simulations and \textit{All of Us} implementation, the proposed \textit{RAILS} estimator has demonstrated clear advantages. Traditional calibration raking methods align weighted non-probability samples with known population characteristics, but they assume equal base weights and rely on a manual search for additional calibration factors. In contrast, the \textit{RAILS} method provides a unified, scalable, and data-supported solution that automates and improves calibration raking. This approach substantially improves calibration accuracy and efficiency by effectively leveraging two types of high-quality external data. With developed sampling weights for \textit{All of Us}, we can estimate the prevalence of the US population for a broad range of genotypes and phenotypes or conduct representative association studies using GWAS and PheWAS, addressing the gaps left by traditional national probability samples and improving public health research and policy.

Note that if the primary objective is to estimate disease prevalence, doubly robust estimators, constructed by integrating probability and non-probability samples, could offer a promising solution (\cite{chen_doubly_2020}, \cite{yang_doubly_2020}). Our \textit{RAILS} method can serve as a foundation for building such estimators.

Variance estimation remains a crucial area for further research. In practice, achieving an unbiased variance estimate is difficult without oracle weights, and the assumption of independent inclusion probabilities is often unrealistic, especially in large recruitment centers. Such dependencies frequently arise in non-probability sampling schemes for hard-to-survey populations, such as those recruited through snowball sampling. Additionally, rising non-response rates and missing data in opt-in designs can lead to substantial variance underestimation, ultimately compromising inference reliability and calling for advancement. 

Despite its strengths, the \textit{RAILS} approach does not fully eliminate the challenges of using non-probability samples with high-order cross-tabulations, as other unmeasured factors could influence volunteers' decision to participate in the non-probability samples. 
One limitation of the LIFO strategy is that its greedy algorithm can lead to local optima. Future work could focus on adaptive variable selection, allowing the workflow to incorporate geographic or temporal subgroup differences. For example, state- or region-specific weighting variables could help identify distinct geographic patterns. Additionally, in the \textit{All of Us} study, approximately 25\% of participants are excluded from the analysis due to the missing demographic information required by the calibration method. Some researchers suggest recalculating the weights and adjusting for the missing percentage of variables used in each calibration step (see \cite{slud2022methodology} for details). However, the performance and underlying assumptions of this approach remain unclear. Extending the proposed method to handle samples with missing data is another important extension to be explored. 

\section*{Funding}\label{sec:fund}
This work was supported by \if1\blind{NIH/NIMHD R21MD019103}\fi.

\bibliographystyle{Chicago}
\bibliography{JSM-2025}

\begin{table}
\begin{center}
\caption{Simulation Results for Estimating Outcome Prevalence (Settings S1-S2)} \label{tab:sim1}
{\scriptsize
\resizebox{0.9\textwidth}{!}{
\begin{tabular}{l l c *{7}{d{5.2}} }
\toprule
\toprule
\mc{Scenario} & \mc{Methods} & \mc{\text{Rel Bias}} & \mc{\text{AVar}} & \mc{EVar} & \mc{Nom-CP} & \mc{Ora-CP} & \mc{Divergent} \\
 & & \mc{$(\%)$} & \mc{($\times 10^5$)} & \mc{($\times 10^5$)} & \mc{($\% $)} & \mc{($\% $)} & \mc{($\% $)} \\
\midrule
S1 & \textit{naive} 
 & 54.16 & 0.10 & 0.28 & 0.0 & 100.0 & 0.0 \\
& \textit{oracle} 
& ~-0.01 & 1.03 & 0.97 & 95.1 & 95.1 & 0.0 \\ \cline{2-8}
& \textit{cal-1} 
& \cellcolor{lightgray}~~1.27 & 1.47 & 1.26 & \cellcolor{lightgray}96.0 & \cellcolor{lightgray}96.3 & 0.0 \\
& \textit{nps-1} 
& 19.60 & 3.38 & 3.23 & 23.9 & \cellcolor{lightgray}95.1 & 0.0 \\
& \textit{nps-cal-1} 
& \cellcolor{lightgray}~1.07 & 1.48 & 1.27 & \cellcolor{lightgray}96.2 & \cellcolor{lightgray}96.1 & 0.0 \\
& \textit{vs-nps} 
& -1.64 & 10.26 & 16.16 & 82.4 & 84.6 &  0.0\\
& \textit{vs-rake} 
& ~1.51 & 1.22 & 0.98 & 97.0 & \cellcolor{lightgray}96.3 & 27.5 \\
& \textit{RAILS} 
& ~2.02 & 1.23 & 1.13 & \cellcolor{lightgray}94.1 & \cellcolor{lightgray}94.4 & 0.1 \\
\midrule
S2 & \textit{naive} 
& -30.07 & 0.10 & 0.27 & 0.0 & 100.0 & 0.0 \\
& \textit{oracle} 
& 0.10 & 6.94 & 6.73 & 94.7 & 94.9 & 0.0 \\ \cline{2-8}
& \textit{cal-1} 
& 1.37 & 6.14 & 5.67 & \cellcolor{lightgray}95.8 & \cellcolor{lightgray}95.4 & 0.0 \\
& \textit{nps-1} 
& 15.59 & 8.81& 12.22 & 23.8 & 90.4 & 0.0 \\
& \textit{nps-cal-1} 
& 1.41 & 6.30 & 5.72 & \cellcolor{lightgray}95.8 & \cellcolor{lightgray}95.4 & 0.0 \\
& \textit{vs-nps} 
& \cellcolor{lightgray}-0.37 & 34.35 & 82.12 & 84.1 & 84.9 & 0.0 \\
& \textit{vs-rake} 
& \cellcolor{lightgray}0.99 & 7.30 & 6.70 & \cellcolor{lightgray}95.9 & \cellcolor{lightgray}95.0 & 10.6 \\
& \textit{RAILS} 
& \cellcolor{lightgray}0.99 & 7.20 & 6.55 & \cellcolor{lightgray}95.8 & \cellcolor{lightgray}95.1 & 1.7 \\
\bottomrule
\bottomrule
\end{tabular} %
}}
\end{center}
\clearpage
Rel Bias is the relative bias; AVar is the average of variance estimates; EVar is empirical variance; Nom-CP is the nominal coverage probability for 95$\%$ CI; Ora-CP is the Oracle coverage probability for 95$\%$ CI. Values with good performance were highlighted.
\end{table}

\begin{table}
\begin{center}
\caption{Simulation Results for Estimating Outcome Prevalence (Settings S3-S5)} \label{tab:sim2}
{\scriptsize
\resizebox{0.9\textwidth}{!}{%
\begin{tabular}{l l c *{7}{d{5.2}} }
\toprule
\toprule
\mc{Scenario} & \mc{Methods} & \mc{\text{Rel Bias}} & \mc{\text{AVar}} & \mc{EVar} & \mc{Nom-CP} & \mc{Ora-CP} & \mc{Divergent} \\
 & & \mc{$(\%)$} & \mc{($\times 10^5$)} & \mc{($\times 10^5$)} & \mc{($\% $)} & \mc{($\% $)} & \mc{($\% $)} \\
\midrule
S3 & \textit{naive} 
& 105.84 & 0.14 & 0.40 & 0.0 & 100.0 & 0.0 \\
& \textit{oracle} 
& ~~0.03 & 0.50 & 0.51 & 94.3 & 99.2 & 0.0 \\ \cline{2-8}
& \textit{cal-1} 
& -13.67 & 0.32 & 0.17 & 0.0 & 99.2 & 0.0 \\
& \textit{nps-1} 
& -23.49 & 0.36 & 1.67 & 0.0 & 65.2 & 0.0 \\
& \textit{nps-cal-1} 
& -13.99 & 0.32 & 0.17 & 0.0 & 99.2 & 0.0 \\
& \textit{vs-nps} 
& ~-8.52 & 1.34 & 10.78 & 39.1 & 64.1 & 0.0\\
& \textit{vs-rake} 
& \cellcolor{lightgray}~-3.39 & 0.47 & 0.13 & \cellcolor{lightgray}88.8 & 100.0 & 27.5 \\
& \textit{RAILS} 
& \cellcolor{lightgray}~-2.83 & 0.47 & 0.35 & \cellcolor{lightgray}89.3 & \cellcolor{lightgray}94.3 & 0.1 \\
\midrule
S4 & \textit{naive} 
& 107.01 & 0.10 & 0.25 & 0.0 & 100.0 & 0.0 \\
& \textit{oracle} 
& ~~0.14 & 0.24 & 0.24 & 94.7 & 94.9 & 0.0 \\ \cline{2-8}
& \textit{cal-1} 
& -15.41 & 0.15 & 0.10 & 0.0 & 98.5 & 0.0 \\
& \textit{nps-1} 
& -23.92 & 0.18 & 0.75 & 0.0 & 66.7 & 0.0 \\
& \textit{nps-cal-1} 
& -15.72 & 0.15 & 0.10 & 0.0 & 98.5 & 0.0 \\
& \textit{vs-nps} 
& -12.34 & 0.34 & 6.74 & 34.1 & 44.3 & 0.0 \\
& \textit{vs-rake} 
& \cellcolor{lightgray} ~-3.67 & 0.22 & 0.13 & \cellcolor{lightgray}80.3 & 97.4 & 6.1 \\
& \textit{RAILS} 
& \cellcolor{lightgray}~~-3.45 & 0.22 & 0.17 & \cellcolor{lightgray}80.9 & \cellcolor{lightgray}95.4 & 0.1 \\
\midrule
S5 & \textit{naive} 
& 107.01 & 0.10 & 0.25 & 0.0 & 100.0 & 0.0 \\
& \textit{oracle} 
& ~~0.14 & 0.24 & 0.24 & 94.7 & 94.9 & 0.0 \\ \cline{2-8}
& \textit{cal-1} 
& -15.41 & 0.15 & 0.10 & 0.0 & 98.5 & 0.0 \\
& \textit{nps-1} 
& -24.30 & 0.18 & 0.78 & 0.0 & 64.1 & 0.0 \\
& \textit{nps-cal-1} 
& -15.73 & 0.15 & 0.10 & 0.0 & 98.5 & 0.0 \\
& \textit{vs-nps} 
& -17.90 & 1.84 & 18.27 & 26.6 & 17.5 & 0.0 \\
& \textit{vs-rake} 
& \cellcolor{lightgray}~-3.72 & 0.22 & 0.11 & \cellcolor{lightgray}81.7 & 98.9 & 10.6 \\
& \textit{RAILS} 
& \cellcolor{lightgray}~-3.44 & 0.22 & 0.22 & \cellcolor{lightgray}81.9 &\cellcolor{lightgray}94.4 & 1.7 \\
\bottomrule
\bottomrule
\end{tabular} %
}}
\end{center}
\clearpage
See Table~\ref{tab:sim1} for notations.
\end{table}

\begin{table}
\begin{center}
\caption{Comparison of Auxiliary Variables Between the US Census 2022 and \textit{All of Us} Research Program (Unweighted and Weighted with \textit{RAILS} Method).}
\label{tab:AoU1}
\resizebox{\textwidth}{!}{%
\begin{tabular}{l l c *{8}{d{4.2}} }
\toprule
\toprule
 \textbf{Auxiliary} & \textbf{Category} & \multicolumn{2}{c}{\textbf{US Census}} & \multicolumn{3}{c}{\textbf{\textit{All of Us (Unweighted)}}} & \multicolumn{3}{c}{\textbf{\textit{All of Us (Weighted)}}}\\
\cmidrule(lr){3-4} \cmidrule(lr){5-7} \cmidrule(lr){8-10}
\cmidrule(lr){3-4} \cmidrule(lr){5-7} \cmidrule(lr){8-10}
\textbf{Variables} & & \mc{\textbf{Size $\times 10^7$}} & \mc{\textbf{Percent $\%$}} & \mc{\textbf{Size $\times 10^3$}}  & \mc{\textbf{Percent $\%$}} & \mc{\textbf{Diff $\%$}}\textsuperscript{b} & \mc{\textbf{Size $\times 10^7$}}  & \mc{\textbf{Percent $\%$}} & \mc{\textbf{Diff $\%$}}\textsuperscript{b} \\ 
\midrule
\textbf{Age group} & 
18--24 & 2.88 & 11.04 & 1.80 & 0.84 & -10.20 & 2.88 & 11.04 & 0.00 \\ 
& 25--44 & 9.05 & 34.65 & 57.28 & 26.83 & -7.82 & 9.05 & 34.65 & 0.00 \\ 
& 45--64 & 8.39 & 32.14 & 73.35 & 34.35 & ~2.21 & 8.39 & 32.14 & 0.00 \\ 
& 65--74 & 3.44 & 13.18 & 46.08 & 21.58 & ~8.40 & 3.44 & 13.18 & 0.00 \\ 
& $\geq$75 & 2.34 & 8.98 & 35.00 & 16.39 & ~7.41 & 2.34 & 8.98 & 0.00 \\
\midrule
\textbf{Sex} 
& Female & 13.36 & 51.19 & 130.98 & 61.34 & ~10.15 & 13.36 & 51.19 & 0.00 \\ 
& Male & 12.74 & 48.81 & 82.54 & 38.66 & -10.15 & 12.74 & 48.81 & 0.00 \\ 
\midrule
\textbf{Race/Ethnicity}
& Hispanic & 4.51 & 17.28 & 32.87 & 15.39 & -1.89 & 4.51 & 17.28 & 0.00 \\ 
& Non-Hispanic Asian & 1.61 & 6.16 & 6.29 & 2.95 & -3.21 & 1.61 & 6.16 & 0.00 \\ 
& Non-Hispanic Black & 3.05 & 11.68 & 39.75 & 18.62 & ~6.94 & 3.05 & 11.68 & 0.00 \\ 
&  Non-Hispanic White & 16.63 & 63.71 & 127.52 & 59.72 & -3.99 & 16.63 & 63.71 & 0.00 \\ 
&  Others & 0.31 & 1.17 & 7.08 & 3.32 & ~2.15 & 0.31 & 1.17 & 0.00 \\ 
\midrule
\textbf{Region}
& Northeast & 4.53 & 17.35 & 66.21 & 31.01 & ~13.66 & 4.53 & 17.35 & 0.00 \\ 
&  Midwest & 5.37 & 20.57 & 54.85 & 25.69 & ~5.12 & 5.37 & 20.57 & 0.00 \\ 
&  South & 10.02 & 38.39 & 35.40 & 16.58 & -21.81 & 10.02 & 38.39 & 0.00 \\ 
&  West & 6.18 & 23.69 & 57.06 & 26.72 & ~3.03 & 6.18 & 23.69 & 0.00 \\
\midrule
\textbf{Income}
& $\leq$35k & 4.60 & 17.61 & 86.92 & 40.71 & ~23.10 & 4.60 & 17.61 & 0.00 \\ 
& 35k-50k & 2.50 & 9.57 & 21.19 & 9.92 & ~0.35 & 2.50 & 9.57 & 0.00 \\ 
&  50k-75k & 4.18 & 16.00 & 28.14 & 13.18 & -2.82 & 4.18 & 16.00 & 0.00 \\ 
&  75k-100k & 3.58 & 13.70 & 21.61 & 10.12 & -3.58 & 3.58 & 13.70 & 0.00 \\ 
& $\geq$100k & 11.25 & 43.11 & 55.66 & 26.07 & -17.04 & 11.25 & 43.11 & 0.00 \\
\midrule
\textbf{Education}
&  Less than highschool & 1.10 & 4.23 & 4.69 & 2.20 & -2.03 & 1.10 & 4.23 & 0.00 \\ 
&  Some highschool & 1.58 & 6.07 & 11.27 & 5.28 & -0.79 & 1.58 & 6.07 & 0.00 \\ 
&  Highschool graduate & 7.02 & 26.90 & 38.37 & 17.97 & -8.93 & 7.02 & 26.90 & 0.00 \\ 
&  Some college & 7.56 & 28.96 & 56.85 & 26.63 & -2.33 & 7.56 & 28.96 & 0.00 \\ 
&  College graduate or advanced & 8.84 & 33.85 & 102.33 & 47.93 & ~14.08 & 8.84 & 33.85 & 0.00 \\
\midrule
\textbf{House Tenure}
& Own & 18.04 & 69.12 & 107.42 & 50.31 & -18.81 & 18.04 & 69.12 & 0.00 \\ 
&  Rent & 7.66 & 29.36 & 84.03 & 39.35 & ~9.99 & 7.66 & 29.36 & 0.00 \\ 
&  Others & 0.40 & 1.52 & 22.07 & 10.34 & ~8.82 & 0.40 & 1.52 & 0.00 \\ 
\midrule
\bottomrule
\end{tabular}%
}
\end{center}
\textsuperscript{a}: weights are provided in PUMS 2022 from the Census Bureau’s American Community Survey Data. \textsuperscript{b}: Diff is the percentage difference between \textit{All of Us} (unweighted or weighted) and PUMS 2022.
\end{table}

\begin{landscape}
\begin{table}
\begin{center}
\caption{Comparison of Health Outcomes Between the National Health Interview Survey (NHIS) 2020 and \textit{All of Us} Research Program.}
\label{tab:AoU2}
\resizebox*{9.3in}{4.8in}{%
\begin{tabular}{l c d{2}@{\hspace{7em}}d{2} *{3}{d{2}@{\hspace{7em}}d{2} d{2}}}
\toprule
\toprule
 \textbf{Outcome} & \multicolumn{2}{c}{\textbf{NHIS}} & \multicolumn{3}{c}{\textbf{\textit{All of Us}}} & \multicolumn{3}{c}{\textbf{\textit{All of Us}}} & \multicolumn{3}{c}{\textbf{\textit{All of Us}}}\\
\cmidrule(lr){2-3} \cmidrule(lr){4-6} \cmidrule(lr){7-9} \cmidrule(lr){10-12} 
\textbf{Variables} & \multicolumn{2}{c}{Weighted \textsuperscript{a}} & \multicolumn{3}{c}{Unweighted}  &  \multicolumn{3}{c}{Recalibrate\textsuperscript{b}} &  \multicolumn{3}{c}{Double-Weighting\textsuperscript{b}} \\
\cmidrule(lr){2-3} \cmidrule(lr){4-6} \cmidrule(lr){7-9} \cmidrule(lr){10-12} 
& \mc{\textbf{Size}} & \mc{\textbf{Percent (95\% CI)}}
& \mc{\textbf{Size}}  & \mc{\textbf{Percent (95\% CI)}} & \mc{\textbf{Diff}}\textsuperscript{c} 
& \mc{\textbf{Size}}  & \mc{\textbf{Percent  (95\% CI)}} &  \mc{\textbf{Diff}}\textsuperscript{c}
& \mc{\textbf{Size}}  & \mc{\textbf{Percent  (95\% CI)}} &  \mc{\textbf{Diff}}\textsuperscript{c} \\ 
& \mc{($\times 10^7$)} & \mc{($\%$)} & \mc{($\times 10^3$)} & \mc{($\% $)} & \mc{($\% $)} & \mc{($\times 10^7$)} & \mc{($\% $)} & \mc{($\% $)} & \mc{($\times 10^7$)} & \mc{($\% $)} & \mc{($\% $)} \\
\midrule
\multicolumn{12}{l}{\textbf{Common Visit Careplace (minimally dynamic)}} \\
Doctors office & 19.86 & 87.95 \,(87.95 - 87.96) & 114.34 & 89.98 \, (89.83 - 90.13) & -2.03 & 23.24 & 89.04\, ( 89.04 - 89.04) & -1.09 & 9.58 & 87.69 \, (87.68 - 87.69) & 0.26 \\ 
  Urgent & 1.70 & 7.53\,(7.52 - 7.54) & 7.02 & 5.52 \, (5.09 -5.99) & 2.01 & 1.93 & 7.38 \, (7.37 - 7.39) & 0.15 & 0.84 & 7.73 \, (7.72 - 7.75) & -0.20 \\ 
  Emergency & 0.37 & 1.64\,(1.63 - 1.65) & 2.50 & 1.97 \, (1.56 - 2.48) & -0.33 & 0.39 & 1.49 \, (1.48 - 1.50) & 0.15 & 0.27 & 2.46 \, (2.45 - 2.48) & -0.82 \\ 
  Others & 0.58 & 2.55\,(2.54 - 2.56) & 1.26 & 0.99 \, (0.62 - 1.57) & 1.56 & 0.22 & 0.83 \,( 0.82 - 0.84) & 1.72 & 0.09 & 0.84 \, (0.82 - 0.85) & 1.71 \\ 
  None & 0.07 & 0.32\,(0.31 - 0.33) & 1.95 & 1.54 \, (1.14 - 2.07) & -1.22 & 0.33 & 1.26 \, (1.25 - 1.27) & -0.94 & 0.14 & 1.28 \, (1.27 - 1.30) & -0.96 \\ 
  \cmidrule(lr){4-6} \cmidrule(lr){7-9} \cmidrule(lr){4-6} \cmidrule(lr){7-9} \cmidrule(lr){10-12} 
& & & \multicolumn{3}{c}{Total Absolute Diff = 7.15\%}& \multicolumn{3}{c}{Total Absolute Diff = 4.05\%} & \multicolumn{3}{c}{Total Absolute Diff = 3.95\%}\\ 
\midrule
\multicolumn{12}{l}{\textbf{Self-Reported Health (moderately dynamic)}} \\
Excellent & 6.13 & 24.32\,(24.31 - 24.33) & 31.30 & 11.07 \, (10.78 - 11.37) & 13.25 & 3.11 & 11.93 \, (11.92 - 11.94) & 12.39 & 3.08 & 11.92 \, (11.91 - 11.93) & 12.40 \\ 
  Very Good & 8.64 & 34.31\,(34.30 - 34.32) & 85.67 & 30.30 \, (30.04 - 30.56) & 4.01 & 9.04 & 34.62 \, (34.62 - 34.63) & -0.31 & 8.94 & 34.61 \, (34.60 - 34.62) & -0.30 \\ 
  Good & 6.93 & 27.52\,(27.51 - 27.53) & 96.96 & 34.29 \, (34.04 - 34.54) & -6.77 & 9.08 & 34.79 \, (34.78 - 34.80) & -7.27 & 8.98 & 34.80 \, (34.79 - 34.81) & -7.28 \\ 
  Fair & 2.72 & 10.82\,(10.81-10.83) & 55.06 & 19.47 \, (19.20 - 19.75) & -8.65 & 3.94 & 15.08 \,( 15.07 - 15.09) & -4.26 & 3.90 & 15.09 \, (15.08 - 15.10) & -4.27 \\ 
  Poor & 0.76 & 3.03\,(3.02 - 3.04) & 13.77 & 4.87 \, (4.58 - 5.18) & -1.84 & 0.93 & 3.58 \, (3.57 - 3.59) & -0.55 & 0.93 & 3.58 \, (3.57 - 3.59) & -0.55 \\ 
  \cmidrule(lr){4-6} \cmidrule(lr){7-9} \cmidrule(lr){4-6} \cmidrule(lr){7-9} \cmidrule(lr){10-12} 
& & & \multicolumn{3}{c}{Total Absolute Diff = 34.52\%}& \multicolumn{3}{c}{Total Absolute Diff = 24.78\%} & \multicolumn{3}{c}{Total Absolute Diff = 24.80\%}\\ 
\midrule
\multicolumn{9}{l}{\textbf{Insurance Coverage (most dynamic)}} \\
Not covered & 2.78 & 11.07\,(11.06 - 11.08) & 16.73 & 6.04 \, (5.74 - 6.35) & 5.03 & 1.25 & 4.77 \, (4.76 - 4.78) & 6.30 & 1.24 & 4.82 \, (4.81 - 4.83) & 6.25 \\ 
  Covered & 22.35 & 88.93\,(88.93 - 88.93) & 260.36 & 93.96 \,( 93.88 - 94.04) & -5.03 & 24.86 & 95.23 \, (95.22 - 95.23) & -6.30 & 24.54 & 95.18 \, (95.17 - 95.18) & -6.25 \\ 
  \cmidrule(lr){4-6} \cmidrule(lr){7-9} \cmidrule(lr){4-6} \cmidrule(lr){7-9} \cmidrule(lr){10-12} 
& & & \multicolumn{3}{c}{Total Absolute Diff = 10.06\%}& \multicolumn{3}{c}{Total Absolute Diff = 12.06\%} & \multicolumn{3}{c}{Total Absolute Diff = 12.50\%}\\ 
\midrule
\bottomrule
\end{tabular}%
}
\end{center}
\textsuperscript{a}: Weights are provided in NHIS 2020 with a known sampling scheme. 
\textsuperscript{b}: Methods are described in Section~\ref{Sec:AoU2}.
\textsuperscript{c}: Diff is the percentage difference between \textit{All of Us} 
(unweighted or weighted) and NHIS 2020.
\end{table}
\end{landscape}

\end{document}